\documentclass[11pt,a4paper]{article}
\usepackage{t1enc}
\usepackage[latin1]{inputenc}
\usepackage[english]{babel}
\normalfont
\usepackage{amsmath}
\usepackage{amsthm}
\usepackage{yfonts}

\usepackage{mathabx}
\usepackage{bbm}
\usepackage{bm}
\usepackage{wasysym}
\usepackage{mathrsfs}
\usepackage{pifont}
\usepackage{hyperref}

\newcommand{\be}[0]{\begin{equation}}
\newcommand{\ee}[0]{\end{equation}}

\numberwithin{equation}{section}

\theoremstyle{plain}

\begin{document}

\vspace*{-1cm}
\thispagestyle{empty}
\vspace*{1.5cm}

\begin{center}
{\Large 
{\bf Global gravitational anomaly cancellation for five-branes}}
\vspace{2.0cm}

{\large Samuel Monnier}
\vspace*{0.5cm}

Institut für Mathematik, 
Universität Zürich,\\
Winterthurerstrasse 190, 8057 Zürich, Switzerland\\
samuel.monnier@gmail.com

\vspace*{2cm}

{\bf Abstract}
\end{center}

We show that the global mixed gauge-gravitational anomaly of the worldvolume theory of the M5-brane vanishes, when the anomaly inflow from the bulk is taken into account. 
This result extends to the type IIA and heterotic $E_8 \times E_8$ five-branes. As a by-product, we provide a definition of the chiral fermionic fields for generic non-spin M5-brane worldvolume and determine the coupling between the self-dual field and the M-theory C-field.

\newpage

\tableofcontents

\section{Introduction and summary}

The constraints coming from the cancellation of anomalies have been determinant in the development of string theory. However it is maybe not widely appreciated that we are still far from having checked the cancellation of all anomalies in the low energy effective field theories arising from string theory and M-theory. For instance, relatively few cancellation checks have been performed for global gravitational anomalies \cite{Witten:1985xe}. The latter are anomalous phases picked by the correlation functions of quantum field theories under large diffeomorphisms of spacetime. They have to vanish in any theory in which the diffeomorphism symmetry is supposed to be gauged, and their cancellation is a direct analogue of the well-known constraint of modular invariance on the string worldsheet. Previous checks of global anomaly cancellation in dimension higher than 2 include type I supergravity \cite{Witten:1985xe,Freed:2000ta}, 11-dimensional supergravity  \cite{Witten:1996md, Freed:2004yc} and type IIB supergravity \cite{Monnier2011a}. In the present paper, we use recent results concerning the global gravitational anomaly of the self-dual field \cite{Monnier2011, Monnier2011a, Monniera} in order to check the cancellation of global mixed gauge-gravitational anomalies in M-theory backgrounds containing M5-branes. The cancellation of local gravitational anomalies in such backgrounds was checked in \cite{Duff:1995wd,Witten:1996hc, Freed:1998tg,Lechner:2001sj}. 

Let us quickly recall some generalities about global anomalies. A more detailed discussion can for instance be found in \cite{Witten:1985xe, Freed:1986zx, Monnier:2012pn, Monniera}. 
Consider a quantum field theory in $d$ dimensions coupled to a background metric and/or gauge field. We will denote the space of background fields by $\mathcal{B}$. The partition function $Z$ of the theory can be pictured as a function over $\mathcal{B}$. The group $\mathcal{G}$ of diffeomorphisms and/or gauge transformations act on $\mathcal{B}$. An anomaly is present if $Z$ is not invariant under this action. In dimension higher than 2, it is in general hard to obtain an expression for $Z$ which is explicit enough to check its invariance directly.

The way around this difficulty consists in seeing $Z$ as the pullback to $\mathcal{B}$ of a section of a line bundle $\mathscr{A}$ over the quotient $\mathcal{F} := \mathcal{B}/\mathcal{G}$. It turns out that $\mathscr{A}$ carries a natural connection $\nabla_{\mathscr{A}}$, whose curvature and holonomies are computable. If all the holonomies of $\nabla_{\mathscr{A}}$ turn out to be trivial, then the connection $\nabla_{\mathscr{A}}$ defines a global trivialization of $\mathscr{A}$, and the pullback of any section of $\mathscr{A}$ yields a $\mathcal{G}$-invariant function over $\mathcal{B}$. The triviality of the holonomies of $\nabla_{\mathscr{A}}$ implies the invariance of $Z$, and hence the absence of anomalies.

The well-known \emph{local} anomaly is directly related to the curvature form of $\nabla_{\mathscr{A}}$. But the cancellation of the local anomaly is no guarantee of the invariance of $Z$: $\mathscr{A}$ might still be a flat line bundle with non-trivial holonomies. This corresponds to a situation where the partition function is invariant under infinitesimal diffeomorphisms, but not necessarily under diffeomorphisms disconnected from the identity. The \emph{global} anomaly is the set of holonomies of $\nabla_{\mathscr{A}}$, and its cancellation does ensure the invariance of $Z$ under the full group of diffeomorphisms.

In the case of the M5-brane worldvolume theory, the anomaly has three sources. Two of them are chiral fields on the worldvolume, namely chiral fermions valued in the normal bundle and a self-dual 2-form field. In addition, there is an anomaly inflow due to the Chern-Simons term in the bulk of M-theory. The global anomalies of chiral fermionic theories have been known since the 1980's \cite{Witten:1985xe,MR861886}. A first step toward the understanding of the global anomaly of the self-dual field was taken in \cite{Witten:1985xe}, but it was understood on arbitrary manifolds much more recently in \cite{Monnier2011a,Monniera}. The contribution of the anomaly inflow is given by the holonomies of the Chern-Simons line bundle \cite{Freed:1992vw} defined by the Chern-Simons M-theory term.

However, before we can apply these known formulas, the worldvolume theory has to be defined precisely. The first issue concerns the coupling of the self-dual field to the C-field. As the M5-brane sources the C-field, the latter is defined only on the complement of the M5-brane worldvolume $M$ in the spacetime $Y$. As a consequence, there is no obvious restriction map from $Y$ to $M$ for the C-field. The C-field can however be restricted to a degree 4 shifted differential cocycle $\check{C}$ on the boundary $\tilde{M}$ of a tubular neighborhood of $M$. It appears that the self-dual field on $M$ couples to $\check{C}_M = \frac{1}{2}\pi_\ast(\check{C} \cup \check{C})$, where $\pi_\ast$ is the pushforward map associated to the 4-sphere bundle $\tilde{M} \rightarrow M$. This definition restricts to the intuitive one when the topology of the M5-brane worldvolume is trivial. We also show that under this map, the quantization of $\check{C}$, shifted  by one-fourth of the first Pontryagin class of $Y$, turns into a quantization of $\check{C}_M$ shifted by the Wu class of $M$. The latter shift is exactly what is needed to make sense of the coupling of the worldvolume self-dual field to the C-field \cite{Witten:1996hc, Witten:1999vg, 2012arXiv1208.1540M, Monniera}. The definition of the chiral fermions is also not completely straightforward, as the worldvolume of the M5-brane is not necessarily spin (Section \ref{SecChirFerm}). After these technical details are straightened out, it is relatively easy to check that the total global anomaly vanishes, using earlier results about the local anomaly \cite{Witten:1996hc}.

A limitation of our computation should be mentioned. In order to obtain practical formulas for global anomalies, we need to extend certain topological data from a seven-dimensional manifold $M_c$ to an eight-dimensional manifold $W$ admitting $M_c$ as its boundary. Such extension problems can have obstructions, which can be computed using cobordism theory. In Appendix \ref{SecCobObs}, we describe the relevant cobordism group $\Omega^{M5}_{12}$. We show that it vanishes modulo torsion, but we are currently unable to compute the full group. If $\Omega^{M5}_{12}$ turned out to be non-zero, the anomaly cancellation check above would show that the holonomies of $\nabla_{\mathscr{A}}$ are trivial only along certain loops in $\mathcal{F}$, namely the ones whose associated mapping torus defines a trivial cobordism class.

The paper is organized as follows. We start by quickly reminding the reader of the modeling of gauge fields by differential cocycles in Section \ref{SecRemDiffCoc}, and of the topology and geometry of the M5-brane in Section \ref{SecM5Geom}. We review the M-theory C-field and define its restriction to the worldvolume of the M5-brane in Section \ref{SecMThCField}. We define the space of background fields $\mathcal{B}$ relevant to the study of the anomalies of the worldvolume theory in Section \ref{SecBGFields}. We then describe two constructions which are crucial to the computation of global anomalies. In Section \ref{SecMapTor}, we explain how to associate a mapping torus endowed with a metric and a C-field to a loop in $\mathcal{B}/\mathcal{G}$, and how to extend it to a manifold bounded by the mapping torus in Section \ref{SecBMan}. In Section \ref{SecGlobAn}, we review in turn the contributions to the global anomaly of the chiral fermions, the self-dual field and the anomaly inflow. We check that these contributions cancel in Section \ref{SecTotAn}. Some technical details appear in the Appendices. 

\section{Geometrical setup}

\subsection{A reminder about differential cocycles}

\label{SecRemDiffCoc}

Differential cocycles provide an accurate way of modeling abelian gauge fields and their higher $p$-form analogues (see for instance \cite{Freed:2006yc}). We review briefly this formalism here in order to set up notations. Denote by $C^p(Y,\mathbbm{R})$ the space of $\mathbbm{R}$-valued cochains of degree $p$ on $Y$ and $Z^p(Y,\mathbbm{R})$ the associated space of cocycles. Let $\hat{\lambda} \in Z^p(Y,\mathbbm{R}/\mathbbm{Z})$. (In the following, a hat will denote a cocycle or a cohomology class and a check will denote a differential cocycle.) A differential cochain of degree $p$ on $Y$ shifted by $\hat{\lambda}$ is an element
\be
\check{A} = (a_{\check{A}}, h_{\check{A}}, \omega_{\check{A}}) \in C^p(Y, \mathbbm{R}) \times C^{p-1}(Y, \mathbbm{R}) \times \Omega^p(Y,\mathbbm{R}) = \check{C}(Y) \;,
\ee
subject to $a_{\check{A}} = \hat{\lambda}$ mod 1.
We define a differential $d$ acting on the space of shifted differential cochains as follows: 
\be
d\check{A} = (da_{\check{A}}, \omega_{\check{A}} - dh_{\check{A}} - a_{\check{A}}, d\omega_{\check{A}}) \;,
\ee
where we are abusing the notation by using $d$ for the differentials on the space of differential cochains, on the space of cochains and on the space of differential forms. One can easily check that $d^2 = 0$ on $\check{C}(Y)$. A differential cocycle is a closed differential cochain with respect to $d$, and we will write $\check{Z}_{\hat{\lambda}}^p(Y)$ for the space of differential cocycles of degree $p$ on $Y$ shifted by $\hat{\lambda}$. 

Given a differential cocycle $\check{A}$, we will call $a_{\check{A}}$ the characteristic class of $\check{A}$, and $\omega_{\check{A}}$ the curvature of $\check{A}$. If a degree $p$ differential cocycle $\check{A}$ is used to model an abelian $p-1$-form gauge field, $\omega_{\check{A}}$ is the field strength of the gauge field, so we will use the terms ``curvature'' and ``field strength'' interchangeably. We refer the reader to \cite{Freed:2006yc} and Section 2.2 of \cite{Monniera} for more explanations about the physical interpretation of (shifted) differential cocycles in terms of abelian gauge fields.

We also define an equivalence relation on the space of (shifted) differential cocycles: any two differential $p$-cocycles are equivalent if they differ by the differential of an unshifted differential $p-1$-cochain with vanishing curvature. In equations, if $\check{B} = (a_{\check{B}}, h_{\check{B}}, 0)$, $a_{\check{B}} = 0$ mod 1, 
\be
\label{EqGaugTransDiffCoc}
\check{A} \simeq \check{A} + d\check{B}
\ee
or explicitly
\be
(a_{\check{A}}, h_{\check{A}}, \omega_{\check{A}}) \simeq (a_{\check{A}} + da_{\check{B}}, h_{\check{A}} - dh_{\check{B}} - a_{\check{B}}, \omega_{\check{A}}) \;.
\ee
Equivalent cocycles represent gauge equivalent gauge fields. The gauge group $\mathcal{G}^p_{\rm gauge}$ acting of degree $p$ (shifted) differential cocycles can be described by the following short exact sequence (see Section 2.2 of \cite{Monniera}):
\be
\label{EqShExSeqGaugGrp}
0 \rightarrow \Omega^{p-1}_{\rm exact}(M) \rightarrow \mathcal{G}^p_{\rm gauge} \rightarrow H^{p-1}(M; \mathbbm{Z}) \rightarrow 0 \;.
\ee
$\Omega^{p-1}_{\rm exact}(M)$ is interpreted as the ``small'' gauge transformations shifting the gauge field by exact forms. Modulo torsion, the group of connected components $H^{p-1}(M; \mathbbm{Z})$ describes homotopy classes of ``large'' gauge transformations shifting the gauge field by closed forms with integral periods.

\subsection{The M5-brane geometry}

\label{SecM5Geom}

We consider an M-theory background containing an M5-brane. An eleven-dimensional oriented spin manifold $Y$ represents the M-theory spacetime. As is familiar in the treatment of anomalies, we will take $Y$ to be compact. In the non-compact case, we would anyway be only interested in compactly supported diffeomorphisms and gauge transformations. We could always modify $Y$ 
outside the support of the transformations in order to turn it into a compact manifold. (This can possibly require adding additional five-brane sources in order to satisfy the Gauss law, if there is a non-trivial flux at the boundary of $Y$.) The variation of the partition function under compactly supported transformations (i.e. the anomaly) is not influenced by such modifications. Inside $Y$, we consider a six-dimensional compact oriented manifold $M$, on which an M5-brane is wrapped.

We write $\mathscr{N} \rightarrow M$ for the normal bundle of $M$ in $Y$. From the fact that both $M$ and $Y$ are orientable and $Y$ is spin, the following relations are satisfied by the Stiefel-Whitney classes of $TM$ and $\mathscr{N}$ \cite{Witten:1999vg}:
\be
\label{EqConstrNormBun}
w_1(\mathscr{N}) = 0 \;, \quad w_2(\mathscr{N}) = w_2(M) \;.
\ee

We endow $Y$ with a Riemannian metric $g_Y$. $g_Y$ induces a metric on $TY|_M \simeq TM \oplus \mathscr{N}$. We also obtain a metric $g_M$ on $M$. Let $N$ be the tubular neighborhood of $M$ of constant radius $\delta$, with $0 < \delta <\!\!\!< 1$. In the following, we will always keep the dependence on $\delta$ implicit. $\tilde{M} := \partial N$ is a 4-sphere bundle over $M$ and we will write $\pi$ for the projection map. $\tilde{M}$ inherits a metric $g_{\tilde{M}}$ from its embedding in $Y$.

\subsection{The restriction of the C-field}

\label{SecMThCField}

We will model the M-theory C-field $\check{C}$ by a degree 4 shifted differential cocycle on $Y\backslash M$: $\check{C} \in \check{Z}_{\hat{\lambda}_Y}^4(Y\backslash M)$. The shift $\hat{\lambda}_Y$ is given by $\frac{1}{4}\hat{p}_1(TY)$ modulo 1, half the pullback of a cocycle generating $H^4(BSpin;\mathbbm{Z})$ via a classifying map for $TY$ \cite{Witten:1996md, Diaconescu:2003bm}. As the notation suggests, $4\check{\lambda}_Y$ is an even integral cocycle representing the first Pontryagin class of $TY$. In particular, this means that the field strength $G := \omega_{\check{C}}$ of $\check{C}$ satisfies the shifted quantization law
\be
G = G' + \frac{1}{4}p_1(TY)
\ee
where $G'$ belong to the group $\Omega^4_\mathbbm{Z}(Y \backslash M)$ of degree $4$ differential forms with integral periods on $Y \backslash M$ and $p_1(TY)$ is the first Pontryagin form of $TY$. The periods of $G$ are half-integral.

There exist different equivalent models for the C-field \cite{Diaconescu:2000wy}, in particular a model involving an $E_8$ gauge bundle \cite{Witten:1996md}, which is useful in heterotic M-theory. A general model combining the features of the $E_8$ and differential cohomology models has also been proposed in \cite{Fiorenza:2012mr}. Nevertheless, the model presented above allows for concrete computations and will suffice for our purpose. 

The M5-brane acts as a magnetic source for the C-field. This implies that the integral of $G$ on a 4-sphere in $Y$ linking $M$ is equal to $1$. As a result, $G$, and hence $\check{C}$, is defined only on $Y \backslash M$. This poses a problem, because we expect the self-dual field on the M5-brane worldvolume to be sourced by the restriction of the C-field to $M$. Defining this restriction is non-trivial.

While the C-field cannot be straightforwardly restricted to $M$, it can be restricted to the 4-sphere bundle $\tilde{M}$. This gives a degree 4 differential cocycle $\check{C}_{\tilde{M}}$ shifted by $\hat{\lambda}_{\tilde{M}} = \frac{1}{4}\hat{p}_1(TY|\tilde{M}) = \frac{1}{4}\hat{p}_1(T\tilde{M})$. We have a long exact sequence in cohomology starting by
\be
\label{EqLongExSeqEul}
0 \rightarrow H^4(M;\mathbbm{Z}) \stackrel{\pi^\ast}{\rightarrow} H^4(\tilde{M};\mathbbm{Z}) \stackrel{\pi_\ast}{\rightarrow} H^0(M;\mathbbm{Z}) \stackrel{\hat{e}(\mathscr{N}) \cup}{\rightarrow} ...
\ee
where $\hat{e}(\mathscr{N})$ is the Euler class and $H^0(M;\mathbbm{Z}) \simeq \mathbbm{Z}$. We show in Appendix \ref{SecVanEulClass} that $\hat{e}(\mathscr{N})$ vanishes, which implies that
\be
\label{EqDecompCohomTildM}
H^4(\tilde{M};\mathbbm{Z}) \simeq H^4(M;\mathbbm{Z}) \oplus \mathbbm{Z}
\ee
non-canonically. If there existed a canonical decomposition \eqref{EqDecompCohomTildM}, we could define unambiguously the cohomology class of the restriction of the C-field to $M$ to be the component on $H^4(M;\mathbbm{Z})$ in \eqref{EqDecompCohomTildM}. There exists such a canonical choice of decomposition if $w_4(\mathscr{N}) = 0$, namely half the Euler class of the vertical tangent bundle of $\tilde{M} \rightarrow M$, but not otherwise. Moreover, there seems to be no way to extend the cohomology class to a canonical differential cocycle, as would be needed to define a canonical restriction of the C-field to $M$.

There is however a less obvious way of defining a restriction of the C-field:
\be
\label{EqDefRestCField}
\check{C}_M := \frac{1}{2} \cdot \pi_\ast(\check{C}_{\tilde{M}} \cup \check{C}_{\tilde{M}}) \;,
\ee
where $\cup$ is the cup product on differential cohomology \cite{hopkins-2005-70} and $\frac{1}{2}$ denotes the division by 2 of each of the components of the shifted differential cocycle $\pi_\ast(\check{C}_{\tilde{M}} \cup \check{C}_{\tilde{M}})$. Because of the division by 2, it is a priori not obvious that this definition is gauge invariant, namely that the differential cohomology class of $\check{C}_M$ depends only on the differential cohomology class of $\check{C}_{\tilde{M}}$. This can however be checked explicitly. Upon a gauge transformation $\check{C}_{\tilde{M}} \rightarrow \check{C}_{\tilde{M}} + \check{B}$ with $\check{B}$ as in \eqref{EqGaugTransDiffCoc}, we have $\check{C}_M \rightarrow \check{C}_M + \pi_\ast(\check{C}_{\tilde{M}} \cup \check{B})$, which is indeed a gauge transformation of $\check{C}_M$. We claim that the self-dual field on the M5-brane couples to $\check{C}_M$, as defined in \eqref{EqDefRestCField}.
 
Suppose that we use the parameterization $\check{C}_{\tilde{M}} = \check{f} + \pi^\ast(\check{C}')$, where $\check{f}$ is an unshifted differential cocycle, defining a particular splitting \eqref{EqDecompCohomTildM}, and $\check{C}'$ is a differential cocycle on $M$. Results of Appendix \ref{SecFunctLiftWuClass} show that $\check{C}'$ is shifted by $\frac{1}{4}\hat{p}_1(TM \oplus \mathscr{N})$. Then we have
\be
\label{EqParCM}
\check{C}_M = \frac{1}{2}\pi_\ast(\check{f} \cup \check{f}) + \check{C}' \;,
\ee
so $\check{C}_M$ depends linearly on $\check{C}'$. If the bundle $\mathscr{N}$ is trivial, then $\tilde{M} = M \times S^4$. We can take $\check{f}$ to be a pullback from $S^4$ and the first term in \eqref{EqParCM} vanishes, showing that we recover what we would intuitively call the restriction of the C-field to $M$ in this case, namely $\check{C}_M = \check{C}'$.

Less trivially, we show in Appendix \ref{SecFunctLiftWuClass} that the fact that $\check{C}_{\tilde{M}}$ is a differential cocycle shifted by $\frac{1}{4}\hat{p}_1(TY|_{\tilde{M}})$ implies that $\check{C}_M$ is a differential cocycle shifted by the Wu class of $M$. This is exactly the shift required to couple $\check{C}_{\tilde{M}}$ consistently to the self-dual field on $M$ \cite{Witten:1996hc}, and implies that the global anomaly formula for the self-dual field derived in \cite{Monniera} can be applied. In contrast, the restriction based on a splitting \eqref{EqDecompCohomTildM} does not produce the correct shift, in addition to failing to be unique.

$\check{C}_M$ acts as a source for the self-dual field. Naively, this means that the degree 3 differential cocycle $\check{H}$ representing the self-dual field trivializes $\check{C}_M$. However, the analog of a Freed-Witten anomaly \cite{Freed:1999vc} makes this statement true only up to an unshifted torsion differential cocycle $\check{S}$ \cite{Witten:1999vg,Diaconescu:2003bm,Belov:2006jd,Monniera}:
\be
\label{EqCFSourcSDF}
d\check{H} = \check{C}_M + \check{S}\;,
\ee
where $\check{S} = (a_{\check{S}}, h_{\check{S}}, 0)$ and {$a_{\check{S}} \in Z^4(M,\mathbbm{Z})$ represents a torsion class in integral cohomology. Let us explain this point in more detail. Recall that the Freed-Witten anomaly forces the restriction of the $B$-field to a non-spin$^c$ submanifold wrapped by a D-brane to have a non-trivial torsion characteristic class. Analogously, in the case of the M5-brane, an anomaly forces $[a_{\check{C}_M}] = [a_{\check{S}}] \in H^4(M,\mathbbm{Z})$, i.e. the characteristic class of the C-field is required to restrict to a possibly non-vanishing torsion class on the worldvolume of the M5-brane. If this constraint is not satisfied, the partition function of the self-dual field vanishes identically. There is no closed expression for $\check{S}$, but a detailed characterization can be found in Section 3.7 of \cite{Monniera}. In components, \eqref{EqCFSourcSDF} is equivalent to
\be
da_{\check{H}} = a_{\check{C}_M} + a_{\check{S}} \;, \quad \omega_{\check{H}} - dh_{\check{H}} - a_{\check{H}} = h_{\check{C}_M} + h_{\check{S}} \;, \quad d\omega_{\check{H}} = G_M \;,
\ee
where we write $G_M$ for the field strength of $\check{C}_M$. 
Throughout this paper, we always assume that $\check{C}_M$ satisfies the constraint above. Remark that a shifted differential cocycle can be exact only if the shift is trivial. But the Wu class of degree $k$ of a manifold of dimension less than $2k$ necessarily vanishes, so $\check{C}_M$ is actually an unshifted differential cocycle. Nevertheless, the fact that $\check{C}_M$ is secretely shifted by the Wu class cannot be ignored, it will play an important role in the computation of the anomaly, where $\check{C}_M$ has to be extended to an eight-dimensional manifold.

\subsection{The space of background fields}

\label{SecBGFields}

We can now describe the space of background fields relevant for the study of the M5-brane gauge and gravitational anomalies. 
 
Let $\mathcal{C} \subset \check{Z}^4_{\hat{\lambda}_Y}(Y \backslash M)$ be the space of differential cocycles on $Y \backslash M$ shifted by $\hat{\lambda}_Y$ such that the integral of their field strength along a 4-sphere linking $M$ is 1. Let $\mathcal{M}$ be the space of Riemannian metrics on $Y$. The space of background fields is
\be
\mathcal{B} = \mathcal{C} \times \mathcal{M} \;.
\ee

$\mathcal{B}$ admits a natural action of the group $\mathcal{G}$ generated by gauge transformations and diffeomorphisms. More precisely, the group of gauge transformations of the C-field is $\mathcal{G}_{\rm gauge}^4(Y)$ (cf. \eqref{EqShExSeqGaugGrp}). 
The group of diffeomorphisms $\mathcal{D}$ is the group of orientation-preserving spin diffeomorphisms of $Y$ preserving $M$. The total symmetry group is
\be
\mathcal{G} = \mathcal{G}_{\rm gauge}^4(Y) \rtimes \mathcal{D} \;, 
\ee
where the action of $\mathcal{D}$ on $\mathcal{G}_{\rm gauge}^4(Y)$ is by pullbacks. We will refer to $\mathcal{G}$ as the group of local transformations and write $\mathcal{F} = \mathcal{B}/\mathcal{G}$ for the quotient space. The five-brane partition function and the bulk Chern-Simons term define line bundles with connections over $\mathcal{F}$. Our task to prove the absence of global anomalies in M-theory backgrounds containing five-branes will be to compute the holonomies of these connections along loops in $\mathcal{F}$, and show that they cancel each other for all loops.

\subsection{Mapping tori}

\label{SecMapTor}

In this section, we show how to associate a 4-sphere bundle over a mapping torus of dimension 7 to a loop $c$ in $\mathcal{F}$, endowed with a metric and a C-field. This construction will be of central importance for the computation of global anomalies. 

Given a diffeomorphism $\phi$ of $Y$, the associated mapping torus $Y_\phi$ is the quotient of $Y \times I$ by the equivalence relation $(y,1) = (\phi(y), 0)$. $Y_\phi$ is a fiber bundle over $S^1$ with fiber $Y$. The diffeomorphism entering the construction of $Y_c$ is a diffeomorphism $\phi$ such that $p(1) = \phi^\ast p(0)$, where $p$ is a path in $\mathcal{B}$ lifting $c$. As $\phi$ is a spin diffeomorphism, $Y_c$ is oriented and spin. Combined with the bounding spin structure on $S^1$, the spin structure on $Y$ induces a spin structure on $Y_c$. The choice of the bounding spin structure avoids extra signs in the global anomaly formula for the fermions, as is discussed above Proposition 5.7 of \cite{Dai:1994kq}.

A point in $\mathcal{B}$ determines a metric and a C-field on the space-time $Y$. The path $p$ therefore determines a metric $g_t$ on the fiber above $t$ of $Y_c$, which glues smoothly at $t = 0 \sim 1$. We obtain a metric $g_{Y_c}$ over the full mapping torus by picking a metric $g_{S^1}$ on $S^1$:
\be
g_{Y_c} = g_t \oplus g_{S^1}/\epsilon \;,
\ee
where we rescaled the metric on $S_1$ by a factor $\epsilon \in \mathbbm{R}_+$. The ``adiabatic limit'' \cite{Witten:1985xe} is the limit $\epsilon \rightarrow 0$, in which the size of the base of $Y_c$ blows up. The quantities of interest for the computations of anomalies become independent of $g_{S^1}$ in the adiabatic limit \cite{MR861886,Monniera}.

As the diffeomorphism preserves $M$, we obtain in $Y_c$ a mapping torus $M_c$ with fiber $M$. Let $N_c$ be the tubular neighborhood of $M_c$ in $Y_c$ with radius $\delta$, and $\tilde{M}_c$ its boundary. $\tilde{M}_c$ is a 4-sphere bundle over $M_c$. It is also a mapping torus with fiber $\tilde{M}$. We can restrict the metric and the C-field on $Y_c$ to $\tilde{M}_c$. 
Let $\mathscr{N}_c$ be the normal bundle of $M_c$ in $Y_c$. The orientability of $Y_c$ and $M_c$, together with the fact that $Y_c$ is spin, imply that $M_c$ and $\mathscr{N}_c$ satisfy \eqref{EqConstrNormBun}. 

We also obtain from the path $p$ a family of degree 4 shifted differential cocycles, i.e. of C-fields, living on the fibers of $Y_c \backslash M_c$. It is possible to extend this family to a degree 4 differential cocycle on $Y_c \backslash M_c$ shifted by $\hat{\lambda}_{Y_c} := \frac{1}{4}\hat{p}_1(Y_c)$, as shown in Lemma 3.3 and Corollary 3.4 of \cite{Monniera}.

Finally, denote by $\check{C}_{\tilde{M}_c}$ the restriction of the C-field on $\tilde{M}_c$. We can define a C-field on $M_c$ by $\check{C}_{M_c} = \frac{1}{2}\pi_\ast(\check{C}_{\tilde{M}_c} \cup \check{C}_{\tilde{M}_c})$, where $\pi: \tilde{M}_c \rightarrow M_c$ is the bundle map. $\check{C}_{M_c}$ is a differential cocycle shifted by the Wu class of $M_c$, as shown in Appendix \ref{SecFunctLiftWuClass}.

\subsection{Bounded manifolds}

\label{SecBMan}

Our computation of the global gravitational anomaly will require that the 4-sphere bundle $\tilde{M}_c \rightarrow M_c$ extends to a 4-sphere bundle $\tilde{W} \rightarrow W$, for $W$ a manifold bounded by $M_c$. We also require that the associated rank 5 vector bundle $\mathscr{N}_W \rightarrow W$ satisfies the same conditions \eqref{EqConstrNormBun} as $\mathscr{N}$. In addition, we require that the Euler class of $\mathscr{N}_W$ vanishes. As explained in Appendix \ref{SecVanEulClass}, this is automatic on manifolds of dimension 6 and 7, but not on manifolds of dimension 8.

As the second Stiefel-Whitney class of $TW \oplus \mathscr{N}_W$ vanishes, this bundle admits a spin structure (i.e. a lift of the associated frame bundle to the simply connected cover of $SO(6) \times SO(5)$). We take the spin structure to extend the spin structure present on the corresponding bundle over the boundary $M_c$.

We pick a metric $g_{\tilde{W}}$ on $\tilde{W}$ extending the metric on $\tilde{M}_c$. $g_{\tilde{W}}$ depends implicitly on the parameters $\delta$ and $\epsilon$. We require that the C-field on $\tilde{M}_c$ extends to $\tilde{W}$ as a differential cocycle of degree 4 $\check{C}_{\tilde{W}}$ shifted by $\hat{\lambda}_{\tilde{W}} := \frac{1}{4}\hat{p}_1(\tilde{W})$. The field strength of $\check{C}_{\tilde{W}}$ automatically integrates to 1 on the 4-sphere fibers of $\tilde{W}$. Remark that such a C-field would not exist if the Euler class of $\mathscr{N}_W$ was non-vanishing, as can be seem from the exact sequence \eqref{EqLongExSeqEul}. As before, we define the restriction of the C-field to $W$ as $\check{C}_W = \frac{1}{2} \pi_\ast(\check{C}_{\tilde{W}} \cup \check{C}_{\tilde{W}})$, where $\pi$ is the bundle map. $\check{C}_W$ is a differential cocycle shifted by the Wu class of $W$. Unlike in lower dimension, the Wu class of $W$ does not vanish in general. We will write $G_{\tilde{W}}$ and $G_W$ for the field strengths of $\check{C}_{\tilde{W}}$ and $\check{C}_W$.

Computing the obstruction to the existence of $\tilde{W}$ is a non-trivial cobordism problem that we discuss in Appendix \ref{SecCobObs}. Should such an obstruction exist, the check of anomaly cancellation performed below would not extend to the loops $c$ such that $\tilde{W}$ does not exist, i.e. to the loops $c$ such that $M_c$ defines a non-trivial class in the cobordism group $\Omega^{M5}_{12}$ defined in Appendix \ref{SecCobObs}.

\section{Global anomalies}
\label{SecGlobAn}

We now review in turn the contributions to the global anomaly of the worldvolume theory of the M5-brane, and check that the sum of these contributions vanishes.

\subsection{Chiral fermions}

\label{SecChirFerm}

If $M$ is spin, $\mathscr{N}$ admits a spin structure as well and the fermions of the worldvolume M5-brane theory are (odd) real sections of $\mathscr{S}_+^{TM} \otimes \mathscr{S}^\mathscr{N}$. $\mathscr{S}_+^{TM}$ is the bundle of positive chirality spinors over $M$ and $\mathscr{S}^\mathscr{N}$ is the spin bundle associated to $\mathscr{N}$. In Lorentz signature, $\mathscr{S}_+^{TM}$ carries a quaternionic structure, because the Clifford algebra is quaternionic in 6 dimensions. $\mathscr{S}^\mathscr{N}$ admits as well a quaternionic structure, because the structure group of $\mathscr{N}$ is $SO(5) \simeq Sp(4)$. The product of these two structures defines a real structure on $\mathscr{S}_+^{TM} \otimes \mathscr{S}^\mathscr{N}$, under the action of which the fermions are invariants. However, in general $M$ is not spin and a more elaborate construction is required to define the fermionic fields. 

Let $\mathscr{S}^{TY}$ be the spinor bundle over $Y$. Over $M$, the structure group of $TY$ factorizes to $SO(6) \times SO(5)$. Let $U \subset M$ be an open set such that $w_2(M)|_U = w_2(\mathscr{N})|_U = 0$, and assume that the lift of the $Spin(11)$ principal bundle associated to the spin structure on $Y$ reduces to a $Spin(6) \times Spin(5)$ principal bundle over $U$. We can always cover $M$ with such sets. Over $U$, we can define spin bundles $\mathscr{S}^{TM}|_U$ and $\mathscr{S}^{\mathscr{N}}|_U$ associated to $TM|_U$ and $\mathscr{N}|_U$. We have 
\be
\label{EqRestSpinBunU}
\mathscr{S}^{TY}|_U = \mathscr{S}^{TM}|_U \otimes \mathscr{S}^{\mathscr{N}}|_U \;.
\ee
$\mathscr{S}^{TY}|_M$ is therefore a well-defined bundle over $M$ that reduces on $U$ to $\mathscr{S}^{TM}|_U \otimes \mathscr{S}^{\mathscr{N}}|_U$.

We now explain how to define a ``chiral'' version $\mathscr{S}^{TY}_+$ of $\mathscr{S}^{TY}|_M$, which reduces to $\mathscr{S}^{TM}_+ \otimes \mathscr{S}^{\mathscr{N}}$ over open sets $U$ of the type discussed above. The chirality element $\Gamma_6$ in the Clifford algebra ${\rm Cl}(6)$ defines a decomposition of $\mathscr{S}^{TM}|_U = \mathscr{S}_+^{TM}|_U \oplus \mathscr{S}_-^{TM}|_U$ into chiral spinors, which induces a decomposition $\mathscr{S}^{TY}|_U = \mathscr{S}^{TY}_+|_U \oplus \mathscr{S}^{TY}_-|_U$. We have to check whether this decomposition can be extended globally over $M$. But this is easily seen to be the case. If $M$ is not spin, the $Spin(6)$-valued transition functions that one would use to define $\mathscr{S}^{TM}$ might fail to satisfy the cocycle condition on triple intersections. This failure can at worse be a sign. As the chirality element is an even element of the Clifford algebra, it is globally well-defined. We therefore have a global decomposition
\be
\label{EqRestSpinBunGlob}
\mathscr{S}^{TY}|_M = \mathscr{S}^{TY}_+ \oplus \mathscr{S}^{TY}_- \;.
\ee

Finally, we have to describe a real structure on $\mathscr{S}^{TY}_+$ reducing to the real structure defined above when $M$ is spin. This real structure needs to exist only when the metric on $M$ has Lorentzian signature. But this is immediate: the Clifford algebra is real in 11 dimensions with Lorentzian signature. This induces a real structure on $\mathscr{S}^{TY}$ which in turn induces a real structure on $\mathscr{S}^{TY}_+$.

The chiral fermions on the worldvolume theory of the M5-brane are therefore (odd) sections of $\mathscr{S}^{TY}_+$. The metric on $TY|_M$ defines a Levi-Civita connection on $TY|_M$, which induces a unique spin connection on $\mathscr{S}^{TY}|_M$. We have an associated Dirac operator
\be
\label{EqDiracOp}
D: \Gamma(\mathscr{S}^{TY}|_M) \rightarrow \Gamma(\mathscr{S}^{TY}|_M) \;,
\ee
where $\Gamma(\mathscr{B})$ indicates the space of sections of a bundle $\mathscr{B}$. $D$ changes the parity of the sections with respect $\Gamma_6$, so we can define the chiral Dirac operator
\be
D_+ := D|_{\Gamma(\mathscr{S}^{TY}_+)}: \mathscr{S}^{TY}_+ \rightarrow \mathscr{S}^{TY}_- \;.
\ee
We have $D = D_+ + D_-$, $D_- = (D_+)^\dagger$. 

Consider now a loop $c$ in the space of background fields $\mathcal{F}$. Following the discussion in Sections \ref{SecBMan} and \ref{SecMapTor}, we have a mapping torus $M_c$ endowed with a rank 5 bundle $\mathscr{N}_c$ whose Stiefel Whitney classes satisfy \eqref{EqConstrNormBun}. $\mathscr{N}_c \rightarrow M_c$ is the boundary of a rank 5 vector bundle $\mathscr{N}_W \rightarrow W$ satisfying \eqref{EqConstrNormBun} as well.  

Recall that $TW \oplus \mathscr{N}_W$ is endowed with a spin structure. We can therefore repeat the arguments above and construct the Dirac operator \eqref{EqDiracOp}. The only construction that does not carry over is the real structure. Indeed, $TW \oplus \mathscr{N}_W$ has dimension 13, so the associated Dirac operator $D_W$ is quaternionic in Euclidean signature. (Another way to see this is to remark that the Clifford bundle of $\mathscr{N}_W$ stills carries a quaternionic structure, but that the Clifford bundle of $TW$ is now endowed with a real structure. The tensor product of these two structures yield a quaternionic structure.) Writing $I_f$ for the index density of $D_W$, the standard formula for the global gravitational anomaly of the chiral fermions on the world-volume of the M5-brane is given by an eta invariant \cite{Witten:1985xe, MR861886, Dai:1994kq}. Using the Atiyah-Patodi-Singer theorem \cite{MR0397797}, we can express the latter using $W$ \cite{Witten:1985xe}:
\be
\label{EqGlobAnFerm}
\frac{1}{2\pi i} \ln {\rm hol}_{\mathscr{A}_f}(c) = \lim_{\epsilon \rightarrow 0} \frac{1}{2}\left({\rm index}(D_W) - \int_W I_f \right) \;.
\ee
In the equation above, $\mathscr{A}_f$ is the anomaly line bundle of the chiral fermions, which is the pfaffian line bundle of the chiral Dirac operator $D_+$, endowed with its Bismut-Freed connection \cite{MR861886}. ${\rm hol}_{\mathscr{A}_f}$ is the holonomy of this connection, and in order to compute it, we need to take the adiabatic limit $\epsilon \rightarrow 0$. Remark that as $D_W$ is quaternionic, its index is necessarily even. Therefore the first term on the right-hand side of \eqref{EqGlobAnFerm} does not contribute to the holonomy. 

The degree 8 component of $I_f$ was computed in \cite{Witten:1996hc} in terms of the characteristic classes of $TW$ and $\mathscr{N}_W$. ($\cite{Witten:1996hc}$ implicitly assumed $W$ to be spin, but as the index density can be computed locally, we can always perform the computation by restricting the bundle to an open set $U$ which is spin.):
\begin{align}
I_f^{[8]} & = \left( \hat{A}(TW) {\rm ch}(\mathscr{N}_W) \right)^{[8]} \notag \\
& = \frac{1}{5760} \big(28 p_1(TW)^2 - 16p_2(TW) + 120p_1(TW)p_1(\mathscr{N}_W) \\
& \quad + 60p_1(\mathscr{N}_W)^2 + 240p_2(\mathscr{N}_W) \big) \;, \notag
\end{align}
where $(.)^{[8]}$ denotes the 8-form component.

\subsection{Self-dual 2-form}

\label{SecSDFieldGlobAn}

We now turn to the self-dual 2-form on the worldvolume of the M5-brane. A formula for the global anomaly of the self-dual field was determined in \cite{Witten:1985xe}, but only in the case when the self-dual field has no zero modes, or equivalently when the middle-degree cohomology of the underlying manifold vanishes. A general formula for the gravitational anomaly was determined in \cite{Monnier2011a}, building on \cite{Monnier2011,Monnier:2010ww}. The formula below, which includes the coupling to a background abelian gauge field, was proven in \cite{Monniera}.

The self-dual field can consistently couple only to a gauge field shifted by the Wu class. Fortunately, as we saw above, this is the case for $\check{C}_M$ and its extension $\check{C}_W$. It was shown in \cite{Monniera} that the global mixed gauge-gravitational anomaly of the self-dual field theory associated to a loop $c$ in $\mathcal{F}$ can be computed on $W$ as follows:
\be
\label{EqGlobAnSD}
\frac{1}{2\pi i} \ln {\rm hol}_{\mathscr{A}_{SD}}(c) = \lim_{\epsilon \rightarrow 0} \frac{1}{8} \int_W \left(L(TW) -  4G_W^2 \right) \;.
\ee
${\rm hol}_{\mathscr{A}_{SD}}$ is the holonomy of the natural connection on the anomaly line bundle $\mathscr{A}_{SD}$ of the self-dual field. $L(TW)$ is the Hirzebruch $L$-genus of $TW$, and $G_W$ is the field strength of $\check{C}_W = \frac{1}{2} \pi_\ast(\check{C}_{\tilde{W}} \cup \check{C}_{\tilde{W}})$.

\subsection{Anomaly inflow}

We now turn to the anomaly inflow from the bulk of the M-theory spacetime. Let us recall briefly the inflow mechanism. The low energy effective action of M-theory on $Y$ contains a Chern-Simons term, given schematically by 
\be
CS_{11} =  2\pi i \int_Y \left(\frac{1}{6} C \wedge G \wedge G - C \wedge I_8 \right) \;,
\ee
where
\be
I_8 = \frac{1}{48} \left(p_2(Y) + \left(\frac{p_1(Y)}{2}\right)^2\right) \;.
\ee
These equations are problematic when the C-field is topologically non-trivial and a more refined formulation can be found in \cite{Diaconescu:2003bm}. Recall that $N$ is a tubular neighborhood of $M$ of radius $\delta$. $Y\backslash N$ is a manifold with boundary $\tilde{M}$, and $CS_{11}$ restricted to $Y\backslash N$ is in general not gauge and diffeomorphism invariant. Therefore, under a transformation in $\mathcal{G}$, there will be a change in the bulk M-theory partition function associated to the variation of $CS_{11}$. This is the anomaly inflow, which, if M-theory is anomaly-free, should cancel the contribution of the chiral fermions and of the self-dual field on the M5-brane worldvolume.

A Chern-Simons functional, when considered on a manifold with boundaries, naturally defines a line bundle with connection on the space of boundary data (see for instance \cite{Freed:1992vw}). In our case, the boundary is $\tilde{M}$ and the boundary data can be taken to be the space of background fields $\mathcal{F}$. The holonomy of the Chern-Simons connection along a loop in the space of boundary data is computed by evaluating the Chern-Simons functional on a mapping torus whose fiber is the boundary. Therefore, the anomaly inflow is given by
\be
\label{EqGlobAnInflow}
\frac{1}{2\pi i} \ln {\rm hol}_{\mathscr{A}_{In}}(c) = \lim_{\epsilon,\delta \rightarrow 0} CS_{11}(M_c, \check{C}_{\tilde{M}_c}, g_{\tilde{M}_c}) \;,
\ee
where we made explicit that the Chern-Simons functional is to be evaluated on $\tilde{M}_c$, endowed with the C-field $\check{C}_{\tilde{M}_c}$ and the metric $g_{\tilde{M}_c}$, in the adiabatic limit.

A simple way of evaluating the right-hand side of \eqref{EqGlobAnInflow} is to use the manifold $\tilde{W}$, whose boundary is $\tilde{M}_c$. The Chern-Simons functional on $\tilde{M}_c$ is then obtained by integrating the corresponding characteristic form on $\tilde{W}$, and we get
\be
\label{EqGlobAnInflow2}
\frac{1}{2\pi i} \ln {\rm hol}_{\mathscr{A}_{In}}(c) = \lim_{\epsilon,\delta \rightarrow 0} \int_{\tilde{W}}  \left(\frac{1}{6} G_{\tilde{W}} \wedge G_{\tilde{W}} \wedge G_{\tilde{W}} - G_{\tilde{W}} \wedge I_8\right) \;,
\ee
where the Pontryagin forms in $I_8$ are the Pontryagin forms of $T\tilde{W}$. 

We now pick an arbitrary decomposition 
\be
\label{EqFSCFOnTW}
G_{\tilde{W}} = f_{\tilde{W}} + \pi^\ast F_W \;,
\ee
where $f_{\tilde{W}}$ integrates to 1 on the 4-sphere fibers of $\tilde{W}$, and $F_W$ is a degree 4 differential form on $W$. Let us first plug the explicit expression \eqref{EqFSCFOnTW} into \eqref{EqGlobAnInflow2}. Leaving aside the limits, the right-hand side of \eqref{EqGlobAnInflow2} reads
\be
\label{EqIntFibSt1}
\int_{W} \pi_\ast  \left(\frac{1}{6} (f_{\tilde{W}} + \pi^\ast F_W)^3 - (f_{\tilde{W}} + \pi^\ast F_W) \wedge I_8\right) \;,
\ee
where we decomposed the integral over $\tilde{W}$ into an integral over the fibers of $W$, written as a push-forward $\pi_\ast$, and an integral over $W$. In order to compute the push-forward, we use the relations 
\be
\pi_\ast( \pi^\ast(x)) = 0 \;, \quad \pi_\ast(y \wedge \pi^\ast(x)) = \pi_\ast(y) \wedge x \;, \quad \pi_\ast(f_{\tilde{W}}) = 1 \;,
\ee
valid for $x \in \Omega^\bullet(W)$ and $y \in \Omega^\bullet(\tilde{W})$. We obtain
\be
\label{EqIntFibSt2}
\int_{W}  \left(\frac{1}{6} \pi_\ast(f_{\tilde{W}}^3) + \frac{1}{2} \pi_\ast(f_{\tilde{W}}^2) \wedge F_W + \frac{1}{2}F_W^2 - \pi_\ast(f_{\tilde{W}} \wedge I_8) - F_W \wedge \pi_\ast(I_8) \right) \;.
\ee
As is explained in Appendix \ref{SecFunctLiftWuClass}, the Pontryagin classes of $T\tilde{W}$ appearing in $I_8$ are pullbacks of the Pontryagin classes of $TW \oplus \mathscr{N}_W$ on $W$. This implies that we can integrate the 4th term explicitly and that the last term vanishes. Using $G_W = \frac{1}{2} \pi_\ast(f_{\tilde{W}}^2) + F_W$, we obtain
\be
\label{EqIntFibSt3}
\int_{W}  \left(\frac{1}{6} \pi_\ast(f_{\tilde{W}}^3) - \frac{1}{8} \pi_\ast(f_{\tilde{W}}^2)^2 + \frac{1}{2}G_W^2 - I_8 \right) \;,
\ee
where now the form $I_8$ is constructed from the Pontryagin classes of $TW \oplus \mathscr{N}_W$.

It remains to evaluate 
\be
\label{EqCharClassFibP2}
\frac{1}{6} \pi_\ast(f_{\tilde{W}}^3) - \frac{1}{8} \pi_\ast(f_{\tilde{W}}^2)^2 \;.
\ee
This is most easily done by choosing $f_{\tilde{W}}$ to be half the Euler form of the vertical tangent bundle $T_V\tilde{W}$. This is a differential form which does not necessarily have integral periods, but which does integrate to 1 over the fibers of $\tilde{W}$. Lemma 2.1 of \cite{Bott1998} shows that the first term is equal to $\frac{1}{24}p_2(\mathscr{N}_W)$ and that the second term vanishes. It is also easy to check that \eqref{EqCharClassFibP2} is invariant under $f_{\tilde{W}} \rightarrow f_{\tilde{W}} + \pi^\ast H$, for $H$ any differential form on $W$. We therefore get for the global anomaly inflow
\be
\label{EqGlobAnInflow3}
\frac{1}{2\pi i} \ln {\rm hol}_{\mathscr{A}_{In}}(c) = \lim_{\epsilon \rightarrow 0}
\int_{W}  \left(\frac{1}{24} p_2(\mathscr{N}_W) + \frac{1}{2}G_W^2 - I_8 \right) \;.
\ee

\subsection{Anomaly cancellation}

\label{SecTotAn}

We can now gather the contributions \eqref{EqGlobAnFerm}, \eqref{EqGlobAnSD} and \eqref{EqGlobAnInflow3} to the global anomaly, due respectively to the fermions, the self-dual field and the inflow. After obvious cancellations, we get
\be
\label{EqGlobTot}
\frac{1}{2\pi i} \ln {\rm hol}_{\mathscr{A}_{Tot}}(c) = \lim_{\epsilon \rightarrow 0} \int_{W}  \left(- \frac{1}{2}I_f + \frac{1}{8}L(TW) + \frac{1}{24}p_2(\mathscr{N}_W) - I_8 \right) \;.
\ee
But it was shown in \cite{Witten:1996hc} that these characteristic forms add to zero. Therefore the global anomaly of the M5-brane vanishes, as long as the anomaly inflow contribution from the bulk is taken into account.

There is one last point to discuss to make the cancellation of global anomalies clear. There is in fact another source of global anomaly in the M-theory bulk, namely the gravitino, whose partition function is defined up to a sign. After a choice of sign is made, it can flip under a large diffeomorphism, denoting the presence of a global anomaly. It was shown in \cite{Witten:1996md, Freed:2004yc} that the global anomaly of the gravitino cancels a global anomaly present in the M-theory Chern-Simons term. How come we did not need to take the gravitino into account in the analysis above? This comes from the fact that the global anomaly in 12 dimensions due to the gravitino is purely gravitational. In our formalism, it would be given by the integral of a 12-dimensional index density on $\tilde{W}$ involving only characteristic forms of the tangent bundle of $\tilde{W}$. But we show in Appendix \ref{SecFunctLiftWuClass} that any such characteristic form is a pullback from $W$. This implies that the integral of the relevant index density on $\tilde{W}$ vanishes and the corresponding global anomaly does not play a role in the anomaly cancellation for five-branes.

We have to emphasize that the anomaly cancellation check described above can be performed only when the manifold $W$ described in Section \ref{SecBMan} exists. The obstruction to its existence is given by the cobordism group $\Omega^{M5}_{12}$ described in Appendix \ref{SecCobObs}. If $\Omega^{M5}_{12}$ does not vanish, the check performed above is valid only for loops in $\mathcal{F}$ such that the associated mapping torus $M_c$, together with its extra structure, corresponds to the trivial class in $\Omega^{M5}_{12}$. We show in Appendix \ref{SecCobObs} that $\Omega^{M5}_{12}$ vanishes modulo torsion, but we are currently unable to compute the full group. This implies that a hypothetical global anomaly could affect only a finite number of homotopy classes of loops.

A better strategy would be to check the anomaly cancellation directly on $M_c$, making the existence of $W$ unnecessary. All of the anomalies above can be expressed on $M_c$ (essentially in terms of eta invariants \cite{Witten:1985xe,MR861886,Diaconescu:2003bm}, as well as of an Arf invariant in the case of the self-dual field \cite{Monnier2011a}). The cancellation of the anomaly requires a rather non-trivial relation between these invariants, which we do not currently know how to check directly. Note that a check of the cancellation of global gravitational anomalies in the bulk of M-theory was performed along these lines in \cite{Freed:2004yc}.

A check of the cancellation of global gravitational anomalies of the M5-brane worldvolume theory on a flat 6-dimensional torus was performed in \cite{Dolan:1998qk}. The anomaly inflow from the bulk was not taken into account in this reference, so it might seem surprising that no anomaly was found. This can actually be traced back to the fact that only flat metrics were considered. As the global anomaly vanishes, the global anomaly of the M5-brane worldvolume theory can be expressed in terms the inflow contribution \eqref{EqGlobAnInflow3}. For diffeomorphisms associated to the elementary generators of the $SL(6,\mathbbm{Z})$ mapping class group of the 6-dimensional torus preserving the flat metric, it is not very difficult to see that one can construct $W$ with a flat metric. \eqref{EqGlobAnInflow3} then obviously vanishes. For more general choices of metrics, however, an anomaly is present.

Let us also mention that the check we performed can be trivially extended to the NS5-branes of type IIA string theory and of the $E_8 \times E_8$ heterotic string. Indeed, a type IIA background on a ten-dimensional spacetime $X$ with NS5-branes wrapped on $M \subset X$ can be lifted to M-theory on $Y = X \times S^1$ with M5-branes wrapping $M \times \{p\} \subset Y$, for some $p \in S^1$. Similarly, an $E_8 \times E_8$ heterotic background on $X$ with NS5-branes wrapping $M$ can be lifted to M-theory on $Y = X \times I$ with M5-branes wrapping $M \times \{p\} \subset Y$, for some $p \in I$. The cancellation of global anomalies for the M5-brane therefore implies the cancellation of global anomalies for the NS5-branes in type IIA and heterotic string theories.

\subsection*{Acknowledgments}

This research was supported in part by SNF Grant No.200020-149150/1.

\appendix

\section{The vanishing of the Euler class of the normal bundle}

\label{SecVanEulClass}

In this appendix, we show that given an oriented manifold $X$ of dimension strictly smaller than 8 endowed with a rank 5 vector bundle $\mathscr{N}$ satisfying \eqref{EqConstrNormBun}, then $\hat{e}(\mathscr{N}) = 0$. Remark first that as the rank of $\mathscr{N}$ is odd, we have $2\hat{e}(\mathscr{N}) = 0$. The vanishing of $\hat{e}(\mathscr{N})$ is equivalent to the vanishing of its image in $H^5(M;\mathbbm{Z}_2)$, which is given by $w_5(\mathscr{N})$. We now show that $w_5(\mathscr{N}) = 0$.

As $w_1(\mathscr{N}) = 0$, we have the relation $w_5(\mathscr{N}) = {\rm Sq}^1(w_4(\mathscr{N}))$, where ${\rm Sq}^1$ is the first Steenrod square operation. Moreover, as the dimension of $X$ is smaller than 8, the fourth Wu class $\nu_4 = w_4(\mathscr{N}) + w_2(\mathscr{N})^2$ vanishes. This implies, using the properties of the Steenrod squares, that
\be
w_5(\mathscr{N}) = Sq^1(w_2(\mathscr{N})^2) = 2w_2 Sq^1(w_2(\mathscr{N})) = 0 \;.
\ee
If $X$ has dimension 8 or more, there is no reason for $w_5(\mathscr{N})$ to vanish.

\section{The shifted quantization of the C-field}

\label{SecFunctLiftWuClass}

We show in this appendix that the restriction of the C-field defined by \eqref{EqDefRestCField} has fluxes satisfying a shifted quantization law. The shift is given by the Wu class of the worldvolume. This shifted quantization law is necessary for the consistency of the coupling of the C-field to the self-dual field on the M5-brane worldvolume, as shown in \cite{Witten:1996hc, Witten:1999vg, Monniera}.

Recall that the Wu class of degree 4 is given in terms of the Stiefel-Whitney classes by $\nu_4(X) = w_4(X) + w_2^2(X)$ on an oriented manifold $X$. An integral cohomology class is an integral lift of $\nu_4$ if its reduction modulo 2 coincides with $\nu_4$. 
Any $\mathbbm{Z}_2$-valued cocycle representing $\nu_4$ can be seen as a $\mathbbm{R}/\mathbbm{Z}$-valued cocycle $\hat{\lambda}$.  We will say that a differential cocycle shifted by $\hat{\lambda}$ is ``shifted by the Wu class''. We refer the reader to Section 3.1 of \cite{Monniera} or to Section 3 of \cite{2012arXiv1208.1540M} for a more detailed review of these notions.

We now show that the M5-brane geometry described in Sections \ref{SecM5Geom} and \ref{SecMThCField} implies that \eqref{EqDefRestCField} is a differential cocycle shifted by the Wu class. Let $X$ be a smooth oriented manifold endowed with a rank 5 vector bundle $\mathscr{N}$ such that
\be
\label{EqAssumTopApb}
w_1(TX) = w_1(\mathscr{N}) = 0 \;, \quad w_2(TX) + w_2(\mathscr{N}) = 0 \;, \quad w_5(\mathscr{N}) = 0\;.
\ee
We assume that $TX \oplus \mathscr{N}$ is endowed with a metric and a spin structure. Let $\tilde{X} \stackrel{\pi}{\rightarrow} X$ be the associated 4-sphere bundle. Characteristic classes over $\tilde{X}$ and $X$ are related. We have the basic relation
\be
T\tilde{X} \oplus \mathbbm{R}_{\tilde{X}} = \pi^\ast(TX \oplus \mathscr{N}) \;,
\ee
where $\mathbbm{R}_{\tilde{X}}$ denotes the trivial line bundle over $\tilde{X}$. Therefore, the (stable) characteristic classes of $T\tilde{X}$ are pullbacks of those of $TX \oplus \mathscr{N}$. In particular, $\tilde{X}$ is orientable and spin.

We also assume that $\tilde{X}$ is endowed with a differential cocycle $\check{C}_{\tilde{X}}$ shifted by $\frac{1}{4}\hat{p}_1(T\tilde{X})$ and integrating to $1$ on the 4-sphere fibers of $\tilde{X}$. 
We define $\check{C}_X := \frac{1}{2}\pi_\ast(\check{C}_{\tilde{X}} \cup \check{C}_{\tilde{X}})$. Let us write
\be
\check{C}_{\tilde{X}} = \check{f} + \pi^\ast(\check{C}')\;,
\ee
where $\check{f}$ is a fixed unshifted differential cocycle, whose characteristic class defines a splitting \eqref{EqDecompCohomTildM}, and $\check{C}'$ is a differential cocycle on $X$ shifted by $\frac{1}{4}\hat{p}_1(T\tilde{X})$. Then $\check{C}_X$ is given explicitly by \eqref{EqParCM}. 
Witten showed in \cite{Witten:1999vg} that if $\hat{f}$ is the characteristic cocycle of $\check{f}$, then
\be
\pi_\ast(\hat{f} \cup \hat{f}) = w_4(\mathscr{N}) \quad {\rm mod} \; 2 \;.
\ee
The vanishing of the Euler class of $\mathscr{N}$ is crucial to derive this result. It was also shown in \cite{Witten:1999vg} that $\pi_\ast(\hat{f} \cup \hat{f}) + \frac{1}{2}\hat{p}_1(TX \oplus \mathscr{N})$ is a lift to integral cohomology of $\nu_4(X)$, the Wu class of degree 4 on $X$. Writing $\hat{\lambda}_X := \frac{1}{2} \pi_\ast(\hat{f} \cup \hat{f}) + \frac{1}{4}\hat{p}_1(TX \oplus \mathscr{N})$, we see that \eqref{EqParCM} implies that $a_{\check{C}_X} = \hat{\lambda}_X$ modulo 1, so $\check{C}_X$ is a differential cocycle shifted by the Wu class.

\section{The cobordism obstruction}

\label{SecCobObs}

\subsection{Description of the cobordism obstruction}

Let us recall that we have a 7-manifold $M_c$ endowed with a rank 5 vector bundle $\mathscr{N}_c$ and a differential cocycle $\check{C}_{\tilde{M}_c}$ on $\tilde{M}_c$, the 4-sphere bundle associated to $\mathscr{N}_c$. The Stiefel-Whitney classes of $M_c$ and $\mathscr{N}_c$ satisfy the constraints \eqref{EqConstrNormBun}, and the Euler class of $\mathscr{N}_c$ vanishes by Appendix \ref{SecVanEulClass}. We want a manifold $W$ with $\partial W = M_c$, endowed with a rank 5 vector bundle $\mathscr{N}_W$ and differential cocycle $\check{C}_{\tilde{W}}$ on $\tilde{W}$ restricting to the corresponding structures on $M_c$. Moreover, we want $\mathscr{N}_W$ to satisfy  \eqref{EqConstrNormBun} and its Euler class to vanish.

We construct the cobordism group using a simple extension of the Pontryagin-Thom construction\footnote{We thank user nsrt on Mathoverflow for suggesting this contruction.}, an accessible account of which can be found in \cite{Miller2001}. We first find a classifying spectrum $\mathcal{C}$ for the stable normal bundle of $M_c$ as well as the topological data we want to extend. 
The cobordism obstruction is then classified by a stable homotopy group of the Thom spectrum $T\mathcal{C}$. 

Let us pick an embedding of $M_c$ into $S^{n}$ for some large $n$, and let $\nu$ be the normal bundle. We embed $\mathscr{N}_c$ as a tubular neighborhood $N_c$ of $M_c$ in $S^n$. Let $\nu'$ be the normal bundle of $N_c$. We have $\nu = \nu' \oplus \mathscr{N}_c$. The second constraint in \eqref{EqConstrNormBun} says that $TM_c \oplus \mathscr{N}_c$ is spin, which is stably equivalent to $\nu'$ being spin. We can therefore classify $\nu'$ by a map $\nu' \rightarrow ESpin(n - 12)$. 

$\mathscr{N}_c$ is classified by a map $c_{\mathscr{N}_c}: M_c \rightarrow BSO(5)$. But not any such map is acceptable, because we want the Euler class of $\mathscr{N}_c$ to vanish. The Euler class can be seen as a map $e: BSO(5) \rightarrow K(\mathbbm{Z};5)$. Vector bundle with vanishing Euler class are classified by the homotopy fiber $F$ of this map. $F$ is a topological space (defined up to homotopy equivalence) fitting in the short exact sequence
\be
\label{EqShExSeqF}
0 \rightarrow F \stackrel{\iota}{\rightarrow} BSO(5) \stackrel{e}{\rightarrow} K(\mathbbm{Z};5) \rightarrow 0 \;.
\ee
We have a universal bundle $EF := \iota^\ast(ESO(5))$ over $F$. Any map $\phi: X \rightarrow F$ determines a bundle $\phi^\ast(EF)$ over $X$ with vanishing Euler class and the homotopy classes of such maps classify rank 5 bundles with vanishing Euler class.

We now turn to the C-field $\check{C}_{\tilde{M}_c}$ on $\tilde{M}_c$. As the Euler class of $\mathscr{N}_W$ vanishes, we can find a class integrating to 1 on the fibers of $\tilde{W}$. The non-trivial part of the extension problem therefore amounts to extending a degree 4 integral cohomology class from $M_c$ to $W$. The integral cohomology class associated to $a_{\check{C}_{M_c}}$ is classified by a map from $M_c$ into $K(\mathbbm{Z},4)$.

We see therefore that the data on $M_c$ admits the classifying space
\be
\mathcal{C} = BSpin(n - 12) \times F \times K(\mathbbm{Z},4) \;.
\ee
Seeing $K(\mathbbm{Z},4)$ as the zero-dimensional bundle over itself and taking the Thom space, we get 
\be
T\mathcal{C} = MSpin(n-12) \wedge TF \wedge K(\mathbbm{Z},4)_+ \;,
\ee
where $MSpin$ is the Thom spectrum of the Spin group, $TF$ is the Thom space of $EF$ and $X_+$ is $X$ with an added disjoint basepoint. If we now apply the Pontryagin-Thom construction, we deduce that the relevant cobordism group is given by the stable homotopy group
\be
\Omega^{M5}_{12} = \lim_{n \rightarrow \infty} \pi_{12+n}(MSpin(n) \wedge TF \wedge K(\mathbbm{Z},4)_+) = \tilde{\Omega}^{\rm Spin}_{12}(TF \wedge K(\mathbbm{Z},4)_+) \;.
\ee
We can simplify this expression further. For any generalized homology theory $E$, $E_\bullet(X) = \tilde{E}_\bullet(X_+) = \tilde{E}_\bullet(X) \oplus E_\bullet(\ast)$, where $\ast$ denotes a point and a tilde indicates the reduced homology group. Setting $E$ to be the generalized cohomology theory corresponding to the spectrum $MSpin \wedge TF$, we can write 
\begin{align}
\Omega^{M5}_{12} = \; & \tilde{E}(K(\mathbbm{Z},4)_+) = \tilde{E}(K(\mathbbm{Z},4)) \oplus E(\ast) = \\
= \; & \tilde{\Omega}^{\rm Spin}_{12}(TF) \oplus \tilde{\Omega}^{\rm Spin}_{12}(TF \wedge K(\mathbbm{Z},4)) \notag
\end{align}
In the following, we will show that $\Omega^{M5}_{12}$ vanishes modulo torsion. $\Omega^{M5}_{12}$ can only have 2-torsion, and we expect that the 2-torsion part should be computable using the Adams spectral sequence following the ideas of Section 2 of \cite{Francisa}. Alternatively, it may be possible to compute it using the Atiyah-Hirzebruch spectral sequence. However the computation looks far from straightforward and we will not attempt to carry it out here.

\subsection{Computation of $\Omega^{M5}_{12}$ modulo torsion}

Over $\mathbbm{R}$, the Atiyah-Hirzebruch spectral sequence implies
\be
\label{EqSimpSpinCobOvR}
\tilde{\Omega}^{\rm Spin}_{p}(X) \otimes \mathbbm{R} = \bigoplus_{q = 0}^p \tilde{H}_q(X; \tilde{\Omega}^{\rm Spin}_{p-q}(\ast) \otimes \mathbbm{R}) \;.
\ee
We now apply this formula with $X = TF$. By the Thom isomorphism, $\tilde{H}_{q}(TF;\mathbbm{R}) \simeq H_{q-5}(F;\mathbbm{R})$. Moreover, $\Omega^{\rm Spin}_p \otimes \mathbbm{R}$ is non-vanishing only for $p = 0$ and $4$ if $p$ is less than 8. We have therefore
\be
\tilde{\Omega}^{\rm Spin}_{12}(TF) \otimes \mathbbm{R} = \bigoplus_{q = 0}^7 H_q(F; \tilde{\Omega}^{\rm Spin}_{p-q} \otimes \mathbbm{R}) = H_3(F; \mathbbm{R}) \oplus H_7(F; \mathbbm{R})
\ee
To compute the real cohomology groups of $F$, we remark that $H_{p}(K(\mathbbm{Z};5);\mathbbm{R})$ vanishes except in degree 5, and $H_3(BSO(5); \mathbbm{R}) = H_7(BSO(5); \mathbbm{R}) = 0$. The long exact sequence in homology associated to the short exact sequence \eqref{EqShExSeqF}, reading
\be
... \rightarrow H_{p+1}(K(\mathbbm{Z};5);\mathbbm{R}) \rightarrow H_p(F; \mathbbm{R}) \rightarrow H_p(BSO(5); \mathbbm{R}) \rightarrow  H_{p}(K(\mathbbm{Z};5);\mathbbm{R}) \rightarrow ... \;, 
\ee
then implies that $H_3(F; \mathbbm{R}) = H_7(F; \mathbbm{R}) = 0$.

A completely similar computation shows that $\tilde{\Omega}^{\rm Spin}_{12}(TF \wedge K(\mathbbm{Z},4)) \otimes \mathbbm{R} = 0$.

{
\small

\providecommand{\href}[2]{#2}\begingroup\raggedright\endgroup

}

\end{document}